\documentclass[a4paper,12pt]{article}

\usepackage{ifpdf}

\newif\ifpdf
\ifx\pdfoutput\undefined
  \pdffalse
\else
  \pdfoutput=1
  \pdftrue
\fi

\RequirePackage{xspace} %
\RequirePackage{subfigure} %
\RequirePackage[centertags]{amsmath} %
\RequirePackage{amssymb}
\RequirePackage{wrapfig} %
\RequirePackage{calc} %
\RequirePackage{ifthen}
\RequirePackage{tabularx} %
\RequirePackage{flafter} %
\RequirePackage{fancyhdr} %

\ifpdf
  \RequirePackage[pdftex]{color}%
  \RequirePackage{colortbl}%
  \RequirePackage{array}%
  \RequirePackage[pdftex]{graphicx}

  \RequirePackage[ pdftex, plainpages = false, pdfpagelabels,
                 pdfpagelayout = useoutlines,
                 bookmarks,
                 breaklinks = true,
                 linktocpage,
                 pagebackref,                      
                 colorlinks = true,
                 linkcolor = blue,
                 urlcolor  = blue,
                 citecolor = blue,
                 anchorcolor = blue,
                 hyperindex = true,
                 hyperfigures
                 ]{hyperref}

\else
  \RequirePackage{color}
  \RequirePackage{colortbl}
   \RequirePackage{array}
  \RequirePackage[dvips]{graphicx}
  \RequirePackage{hyperref}
  \usepackage{rotating}
\fi

\usepackage{makeidx} 
\usepackage{setspace} 
\usepackage{rotating} 
\usepackage{ecltree}
\usepackage{epic}
\usepackage{supertabular}  
\usepackage{color}
\usepackage{exscale}
\usepackage{fontenc}
\usepackage{ifthen}
\usepackage{latexsym}
\usepackage{makeidx}
\usepackage{syntonly}
\usepackage{inputenc}
\usepackage{graphicx}
\usepackage{setspace}
\usepackage{caption2}
\usepackage[english]{babel}
\usepackage[square, comma,numbers,sort&compress]{natbib}
\usepackage{hypernat}
\usepackage{boxedminipage}
\usepackage{framed}
\usepackage{longtable}
\usepackage[all]{hypcap}    
\usepackage{algorithm2e}
\usepackage{algorithmic}
\usepackage{lscape}
\usepackage{pdflscape}
\usepackage[T1]{fontenc}
\usepackage{microtype}

\newcommand{\note}[1]{\marginpar[left]{\singlespace \tiny #1}}

\renewcommand{\sectionmark}[1]%
      {\markright{\thesection\ #1}} 

\renewcommand{\note}[1]{}

\newcommand{\etal}{{\it et al}}

\onehalfspace 

\DisableLigatures[f]{encoding = *, family = * }

\begin{document}
\begin{center}
{\Large Variational approach for resolving the flow of generalized Newtonian fluids in circular
pipes and plane slits}
\par\end{center}{\Large \par}

\begin{center}
Taha Sochi
\par\end{center}

\begin{center}
{\scriptsize University College London, Department of Physics \& Astronomy, Gower Street, London,
WC1E 6BT \\ Email: t.sochi@ucl.ac.uk.}
\par\end{center}

\begin{abstract}
\noindent In this paper, we use a generic and general variational method to obtain solutions to the
flow of generalized Newtonian fluids through circular pipes and plane slits. The new method is not
based on the use of the Euler-Lagrange variational principle and hence it is totally independent of
our previous approach which is based on this principle. Instead, the method applies a very generic
and general optimization approach which can be justified by the Dirichlet principle although this
is not the only possible theoretical justification. The results that were obtained from the new
method using nine types of fluid are in total agreement, within certain restrictions, with the
results obtained from the traditional methods of fluid mechanics as well as the results obtained
from the previous variational approach. In addition to being a useful method in its own for
resolving the flow field in circular pipes and plane slits, the new variational method lends more
support to the old variational method as well as for the use of variational principles in general
to resolve the flow of generalized Newtonian fluids and obtain all the quantities of the flow field
which include shear stress, local viscosity, rate of strain, speed profile and volumetric flow
rate. The theoretical basis of the new variational method, which rests on the use of the Dirichlet
principle, also provides theoretical support to the former variational method.

\vspace{0.3cm}


\par\end{abstract}

\begin{center}

\par\end{center}

\clearpage
\section{Introduction} \label{Introduction}

The flow through circular pipes and plane slits has many applications in physical and biological
sciences and engineering and hence it has been investigated in the past by many researchers (e.g.
\cite{Willis1965, Aksel1988, Mannheimer1991, BillinghamF1993, EscudierP1996, Steller2001,
Oliveira2002, BarthBC2008, ChenDZY2009, Kalyon2010, Ellahi2013, ShaikhM2014}) using various methods
of fluid dynamics. Recently we proposed the use of Euler-Lagrange variational principle
\cite{SochiVariational2013} to resolve the flow of generalized Newtonian fluids through circular
pipes. The method is based on minimizing the total stress in the flow conduit in the sense of
minimizing the stress profile in the velocity-varying dimension. This attempt was later expanded
and supported by other investigations \cite{SochiSlitPaper2014, SochiVarNonNewt2014,
SochiCarreauCross2015} where the method was successfully applied to more types of fluid and another
type of geometry, namely the plane slit conduit.

Despite the success of this method in describing the flow of several fluid models and conduit
types, it has not been proven in general by a formal mathematical argument that justifies the
universal applicability of the variational method and the principle on which it relies. Certain
mathematical technicalities may also be disputed and hence it is desirable to fortify the method by
a more generic and general variational approach that is more safe from such disputes and pitfalls.
The present investigation tries to do so by using a very basic and general variational approach
where the flow field in the conduit is resolved through the application of a generic optimization
technique to a stress functional. The theoretical justification of this functional can be obtained
from the Dirichlet principle although it can also be justified by other theoretical foundations
based on purely physical arguments.

The plan for this paper is that in section \S\ \ref{Method} we present a general description of the
proposed method and its theoretical background. This is followed in section \S\
\ref{Implementation} by discussing practical issues about the implementation of this method and the
presentation of sample results that were obtained from this method with comparison to similar
results obtained from the previous methods which include the classical methods of fluid mechanics
and the former variational approach. The paper is finalized in section \S\ \ref{Conclusions} with
general discussions and summarization of the main achievements of the present study. As a matter of
convenience, we label the former variational method which is based on the Euler-Lagrange principle
with EL and the new variational method which is based on the Dirichlet principle with DM.

\clearpage
\section{Method} \label{Method}

\begin{figure}[!t]
\centering{}
\includegraphics
[scale=0.65] {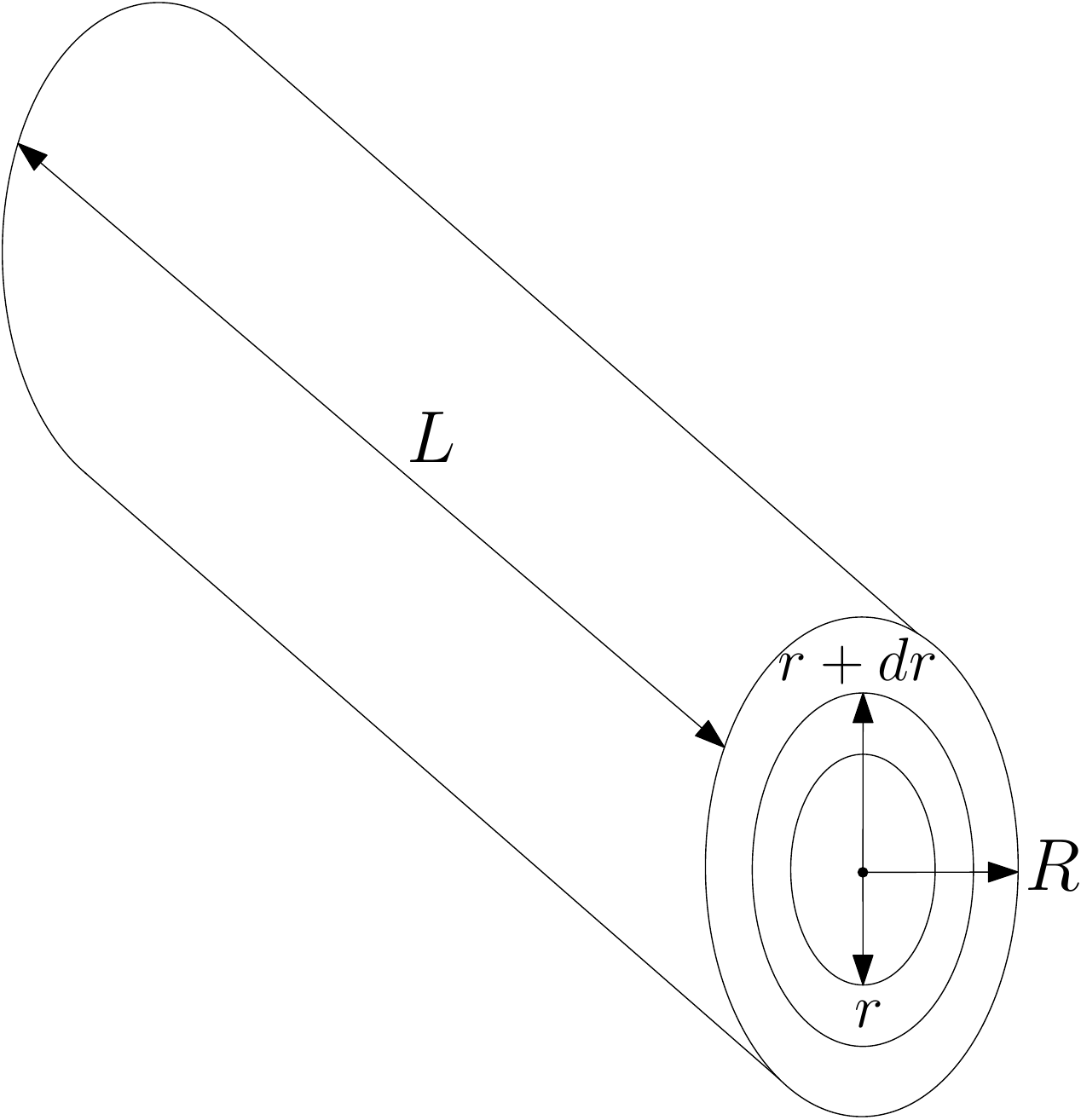} \caption{A schematic of the circular pipe geometry used in this
investigation.} \label{PipePlot}
\end{figure}

\begin{figure}[!h]
\centering{}
\includegraphics
[scale=0.75] {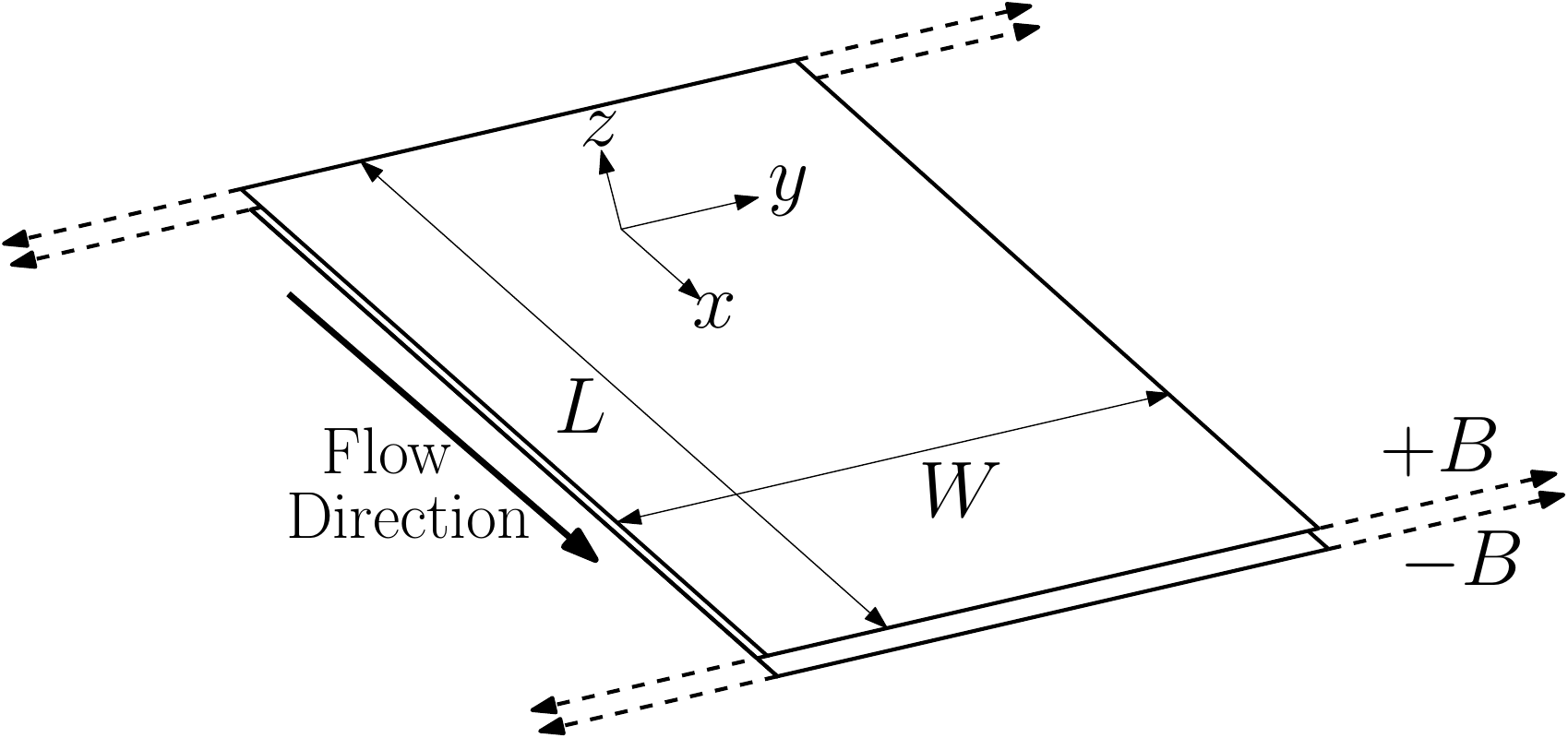} \caption{A schematic of the plane slit geometry used in this
investigation.} \label{SlitPlot}
\end{figure}

We first state our assumptions about the flow, fluid and conduit which are adopted in the present
study. We assume a laminar, isothermal, incompressible, steady, pressure-driven, fully-developed
flow of a time-independent, purely-viscous fluid that can be described by the generalized Newtonian
fluid model, that is
\begin{equation}\label{GNF}
    \tau = \mu \gamma
\end{equation}
where the viscosity, $\mu$, and stress, $\tau$, depend only on the contemporary rate of strain,
$\gamma$, and hence the fluid has no memory of its deformation past. In this formulation we ignore
all non deformation-related dependencies of the viscosity and stress due to other physical factors
like temperature and pressure. In fact we consider only the shearing effects since the effects of
other forms of deformation, such as extensional, are presumed insignificant which is well justified
for the presumed state of flow, fluid and conduit. Edge effects at the entry and exit of the
conduit, as well as external body forces, are also regarded negligible.

Concerning the conduit, we use circular pipe and plane slit geometries, which are depicted in
Figures \ref{PipePlot} and \ref{SlitPlot}, where the pipe is assumed straight with a cross section
that is uniform in shape and size while the slit is assumed straight long and thin with a uniform
cross section. In both cases we assume rigid mechanical characteristics of the conduit wall as
opposite to being deformable such as having elastic or viscoelastic mechanical properties. It is
also assumed that the slit is positioned symmetrically in its thickness dimension, $z$, with
respect to the plane $z=0$ as depicted in Figure \ref{SlitPlot}.

Regarding the boundary conditions, we assume no-slip at the conduit wall, where the fluid
interfaces the solid \citep{SochiSlip2011}, with the flow speed profile having a stationary
derivative point at the symmetry center line of the pipe and symmetry center plane of the slit
which means zero stress and rate of strain at these loci. As for the viscoplastic fluids, this
stationary region expands to include all the points at the forefront of the flow profile whose
stress falls below the yield stress, as will be discussed further in the coming sections.

Now, for the generalized Newtonian fluids that satisfy the above assumptions, the momentum equation
in one dimension is reduced to

\begin{equation}
\frac{d\tau}{ds}=G\label{momEq}
\end{equation}
where $s$ is a spatial coordinate that represents $r$ for pipes and $z$ for slits, and $G$ is a
constant. If we differentiate the last equation with respect to $s$ we get

\begin{equation}
\frac{d^{2}\tau}{ds^{2}}=0
\end{equation}
which is a one-variable Laplace equation in one dimension. According to the Dirichlet principle,
the solution of this equation is a minimizer of the following functional and vice versa

\begin{equation}\label{Dirichlet}
D=\int_{\Omega}\left|\left(\frac{d\tau}{ds}\right)^{2}ds\right|
\end{equation}
where $D$ is the Dirichlet functional and $\Omega$ is a spatial domain in this formulation. On
using Equation \ref{momEq}, substituting and simplifying, the functional can be reduced to

\begin{equation}
D=|G|\int_{T}d\tau
\end{equation}
where $T$ is the new domain in this formulation. The last equation demonstrates that the solution
of this problem is a minimizer (and vice versa) of the total stress in the sense that have been
given previously and hence it establishes the foundation of our former variational approach, EL,
which is based on the use of the Euler-Lagrange principle. It can also provide a theoretical
foundation for new method or methods.

Apart from its theoretical aspects, the Dirichlet principle can be used practically as a basis for
another variational method, DM, that can be employed to obtain flow solutions and verify the
solutions obtained by other methods including the EL method. The DM method in practical terms is
based on finding the shear stress solution in conduits by minimizing the above functional. This
functional is discretized and minimized numerically using an optimization algorithm subject to the
boundary conditions at the conduit wall and conduit center. For pipes these boundary conditions are
respectively

\begin{equation}\label{BCP}
\tau_{w}\equiv \tau_{R}=\frac{R\Delta p}{2L}  \hspace{2cm} {\rm and}  \hspace{2cm} \tau_{m}=0
\end{equation}
while for slits they are

\begin{equation}\label{BCS}
\tau_{w}\equiv \tau_{B}=\frac{B\Delta p}{L}  \hspace{2cm} {\rm and}  \hspace{2cm} \tau_{m}=0
\end{equation}
where $\tau_w$ is the shear stress at the conduit wall which is equivalent to $\tau_R$ for pipes
and to $\tau_B$ for slits, $\tau_m$ is the stress at the conduit center line or plane, $R$ is the
pipe radius, $B$ is the slit half thickness, and $L$ is the conduit length across which a pressure
drop $\Delta p$ is exerted.

The numerically obtained solution, $\tau(s)$, is then used in conjunction with the rheological
constitutive relations, as given in Table \ref{GTable} for the models considered in this study,
applied to the generalized Newtonian fluid equation to find $\gamma(s)$ either explicitly or
implicitly through the use of a simple numerical solver like a bisection solver. As for the
viscoplastic fluids, a zero stress is applied to all the points at the forefront of the flow
profile whose shear stress falls below the yield stress value during the optimization process. As
indicated earlier, this is an extension to the second boundary condition at the conduit center to
include a plane region in the forefront of the speed profile and hence it does not compromise the
optimization condition and the underlying variational principle.

The obtained $\gamma(s)$ is then integrated numerically with respect to $s$ to find the flow speed,
$v(s)$, where the no-slip boundary condition at the conduit wall is used to provide an initial
value, $v=0$, that is incremented on moving inward from the conduit wall toward the conduit center
during the integration process. This is followed by integrating $v(s)$ numerically with respect to
the conduit cross sectional area normal to the flow direction to obtain the volumetric flow rate.
During these successive integration processes, the boundary conditions at the conduit wall and
center, which are based on the zero speed and zero stress respectively, are used.

In Table \ref{GTable} the rheological constitutive relations for the nine fluid models which are
employed in this study are presented, while in Tables \ref{QpTable} and \ref{QsTable} the
analytical relations that correlate the flow rate, $Q$, to the applied pressure drop, $\Delta p$,
for the flow in pipes and slits respectively are given. Most of these expressions can be found in
the classic literature of rheology and fluid dynamics (e.g. \cite{Skellandbook1967,
BirdbookAH1987}) while the rest can be obtained with their derivation from
\cite{SochiVarNonNewt2014, SochiSlitPaper2014, SochiCarreauCross2015}. Regarding the Carreau and
Cross fluids, the ``$I$'' factors, which are included in their $Q$ expressions and represent
definite integral expressions, are given by

\begin{eqnarray}
I_{\rm p,Ca} & = & \frac{\delta^{3}\left[3\lambda^{4}\left(3n'^{2}+5n'+2\right)\gamma_{R}^{4}-3n'\lambda^{2}\gamma_{R}^{2}+2\right]\left(1+\lambda^{2}\gamma_{R}^{2}\right)^{3n'/2}}{3\lambda^{4}\left(9n'^{2}+18n'+8\right)} \nonumber \\
 & + & \frac{\mu_{i}\delta^{2}\left[\lambda^{4}\left(2n'^{2}+5n'+3\right)\gamma_{R}^{4}-n'\lambda^{2}\gamma_{R}^{2}+1\right]\left(1+\lambda^{2}\gamma_{R}^{2}\right)^{n'}}{2\lambda^{4}\left(n'+1\right)\left(n'+2\right)} \nonumber \\
 & + & \frac{\mu_{i}^{2}\delta\left[\lambda^{4}\left(n'^{2}+5n'+6\right)\gamma_{R}^{4}-n'\lambda^{2}\gamma_{R}^{2}+2\right]\left(1+\lambda^{2}\gamma_{R}^{2}\right)^{n'/2}}{\lambda^{4}\left(n'+2\right)\left(n'+4\right)}+\frac{\mu_{i}^{3}\gamma_{R}^{4}}{4} \nonumber \\
 & - &
 \left(\frac{2\delta^{3}}{3\lambda^{4}\left(9n'^{2}+18n'+8\right)}+\frac{\mu_{i}\delta^{2}}{2\lambda^{4}\left(n'+1\right)\left(n'+2\right)}+\frac{2\mu_{i}^{2}\delta}{\lambda^{4}\left(n'+2\right)\left(n'+4\right)}\right)\hspace{1cm}
\end{eqnarray}

{\small
\begin{eqnarray}
I_{\rm p,Cr} & = & \frac{\left\{ 2\delta^{3}\left[-m\left(2f^{2}+5f+3\right)+4g^{2}+2m^{2}\right]+12m\delta^{2}\mu_{i}g\left(m-g\right)+12m^{2}\delta\mu_{i}^{2}g^{2}+3m^{2}\mu_{i}^{3}g^{3}\right\} \gamma_{R}^{4}}{12m^{2}g^{3}} \nonumber \\
 & - & \frac{\left\{ \delta^{3}\left(m^{2}-6m+8\right)+3m\delta^{2}\mu_{i}\left(m-4\right)+3m^{2}\delta\mu_{i}^{2}\right\} {}_{2}F_{1}\left(1,\frac{4}{m};\frac{m+4}{m};-f\right)\gamma_{R}^{4}}{12m^{2}}
\end{eqnarray}
}

\begin{eqnarray}
I_{\rm s,Ca} & = & \frac{n'\delta^{2}\gamma_{B}\left[_{2}F_{1}\left(\frac{1}{2},1-n';\frac{3}{2};-\lambda^{2}\gamma_{B}^{2}\right)-{}_{2}F_{1}\left(\frac{1}{2},-n';\frac{3}{2};-\lambda^{2}\gamma_{B}^{2}\right)\right]}{\lambda^{2}} \nonumber \\
 & + & \frac{\left(1+n'\right)\delta^{2}\gamma_{B}^{3}\,{}_{2}F_{1}\left(\frac{3}{2},-n';\frac{5}{2};-\lambda^{2}\gamma_{B}^{2}\right)}{3} \nonumber \\
 & + & \frac{n'\delta\mu_{i}\gamma_{B}\left[_{2}F_{1}\left(\frac{1}{2},1-\frac{n'}{2};\frac{3}{2};-\lambda^{2}\gamma_{B}^{2}\right)-{}_{2}F_{1}\left(\frac{1}{2},-\frac{n'}{2};\frac{3}{2};-\lambda^{2}\gamma_{B}^{2}\right)\right]}{\lambda^{2}} \nonumber \\
 & + & \frac{\left(2+n'\right)\delta\mu_{i}\gamma_{B}^{3}\,{}_{2}F_{1}\left(\frac{3}{2},-\frac{n'}{2};\frac{5}{2};-\lambda^{2}\gamma_{B}^{2}\right)+\mu_{i}^{2}\gamma_{B}^{3}}{3}
\end{eqnarray}
and

\begin{equation}
I_{\rm s,Cr}=\frac{\left[3\delta^{2}\left(m-g\right)-\left\{
\delta^{2}\left(m-3\right)+2m\delta\mu_{i}\right\}
g^{2}\,_{2}F_{1}\left(1,\frac{3}{m};1+\frac{3}{m};-f\right)+6m\delta\mu_{i}g+2m\mu_{i}^{2}g^{2}\right]\gamma_{B}^{3}}{6mg^{2}}
\end{equation}
where, in these expressions,

\begin{equation}\label{Conditions}
\delta=\left(\mu_{0}-\mu_{i}\right),  \hspace{1cm}  n'=\left(n-1\right),  \hspace{1cm}
f=\lambda^{m}\gamma_{w}^{m},   \hspace{1cm}  g=1+f,
\end{equation}
and $_{2}F_{1}$ is the hypergeometric function of the given arguments with its real part being used
in the evaluation of $I$ factors. Moreover

\begin{equation}
\mu_{R}\gamma_{R}=\tau_{R}  \hspace{2cm} {\rm and}  \hspace{2cm} \mu_{B}\gamma_{B}=\tau_{B}
\end{equation}
where $\tau_{R}$ and $\tau_{B}$ are given by Equations \ref{BCP} and \ref{BCS} respectively, with
$\gamma_{R}$ and $\gamma_{B}$ being obtained numerically from the above implicit relations, as
explained in \cite{SochiCarreauCross2015}.


\renewcommand{\arraystretch}{1.6}

\begin{table} [!h]
\caption{The constitutive relations for the nine fluid models used in this investigation. The
meaning of the symbols are given in Nomenclature \S\ \ref{Nomenclature}. \label{GTable}}
\begin{center} 
{
\begin{tabular}{|l|l|}
 \hline
 Model & \hspace{1cm} Constitutive Relation \\
 \hline
 Newtonian & $\tau=\mu_{o}\gamma$ \\
 Power Law & $\tau=k\gamma^{n}$ \\
 Ellis & $\mu=\mu_{e}\left[1+\left(\frac{\tau}{\tau_{h}}\right)^{\alpha-1}\right]^{-1}$ \\
 Ree-Eyring & $\tau=\tau_{c}\,\mathrm{arcsinh}\left(\frac{\mu_{r}\gamma}{\tau_{c}}\right)$ \\
 Carreau & $\mu=\mu_{i}+\left(\mu_{0}-\mu_{i}\right)\left(1+\lambda^{2}\gamma^{2}\right)^{\left(n-1\right)/2}$ \\
 Cross & $\mu=\mu_{i}+\frac{\mu_{0}-\mu_{i}}{1+\lambda^{m}\gamma^{m}}$ \\
 Bingham & $\tau=C'\gamma+\tau_{0}$ \\
 Herschel-Bulkley & $\tau=C\gamma^{n}+\tau_{0}$ \\
 Casson & $\tau^{1/2}=\left(K\gamma\right)^{1/2}+\tau_{0}^{1/2}$ \\
 \hline
\end{tabular}
}
\end{center}
\end{table}


\begin{table} [!h]
\caption{The volumetric flow rate, $Q$, for the pipe flow of the nine fluid models used in this
investigation. The symbols are given in Nomenclature \S\ \ref{Nomenclature}. \label{QpTable}}
\begin{center} 
{
\begin{tabular}{|l|l|}
 \hline
 Model & \hspace{3cm} $Q$ \tabularnewline
 \hline
 Newtonian & $\frac{\pi R^{4}\Delta p}{8L\mu_{o}}$\tabularnewline
 Power Law & $\frac{\pi R^{4}}{8L}\sqrt[n]{\frac{\Delta p}{k}}\left(\frac{4n}{3n+1}\right)\left(\frac{2L}{R}\right)^{1-1/n}$\tabularnewline
 Ellis & $\frac{\pi R^{3}\tau_{R}}{4\mu_{e}}\left[1+\frac{4}{\alpha+3}\left(\frac{\tau_{R}}{\tau_{h}}\right)^{\alpha-1}\right]$\tabularnewline
 Ree-Eyring & $\frac{\pi R^{3}\tau_{c}}{\tau_{R}^{3}\mu_{r}}\left[\left(\tau_{c}\tau_{R}^{2}+2\tau_{c}^{3}\right)\cosh\left(\frac{\tau_{R}}{\tau_{c}}\right)-2\tau_{c}^{2}\tau_{R}\sinh\left(\frac{\tau_{R}}{\tau_{c}}\right)-2\tau_{c}^{3}\right]$\tabularnewline
 Carreau & $\frac{\pi R^{3}I_{\rm p,Ca}}{\tau_{R}^{3}}$\tabularnewline
 Cross & $\frac{\pi R^{3}I_{\rm p,Cr}}{\tau_{R}^{3}}$\tabularnewline
 Bingham & $\frac{\pi R^{4}\Delta p}{8LC'}\left[\frac{1}{3}\left(\frac{\tau_{0}}{\tau_{R}}\right)^{4}-\frac{4}{3}\left(\frac{\tau_{0}}{\tau_{R}}\right)+1\right]$\tabularnewline
 Herschel-Bulkley & $\frac{8\pi}{\sqrt[n]{C}}\left(\frac{L}{\Delta p}\right)^{3}\left(\tau_{R}-\tau_{0}\right)^{1+1/n}\left[\frac{\left(\tau_{R}-\tau_{0}\right)^{2}}{3+1/n}+\frac{2\tau_{0}\left(\tau_{R}-\tau_{0}\right)}{2+1/n}+\frac{\tau_{0}^{2}}{1+1/n}\right]$\tabularnewline
 Casson & $\frac{\pi R^{3}}{\tau_{R}^{3}K}\left(\frac{\tau_{R}^{4}}{4}-\frac{4\sqrt{\tau_{0}}\tau_{R}^{7/2}}{7}+\frac{\tau_{0}\tau_{R}^{3}}{3}\right)$\tabularnewline
 \hline
\end{tabular}
}
\end{center}
\end{table}


\begin{table} [!h]
\caption{The volumetric flow rate, $Q$, for the slit flow of the nine fluid models used in this
investigation. The symbols are given in Nomenclature \S\ \ref{Nomenclature}. \label{QsTable}}
\begin{center} 
{
\begin{tabular}{|l|l|}
 \hline
 Model & \hspace{3cm} $Q$ \tabularnewline
 \hline
 Newtonian & $\frac{2WB^{3}\Delta p}{3\mu_{o}L}$\tabularnewline
 Power Law & $\frac{2WB^{2}n}{2n+1}\sqrt[n]{\frac{B\Delta p}{kL}}$\tabularnewline
 Ellis & $\frac{2WB^{2}}{\mu_{e}}\left[\frac{\tau_{B}}{3}+\frac{\tau_{B}^{\alpha}}{\left(\alpha+2\right)\tau_{h}^{\alpha-1}}\right]$\tabularnewline
 Ree-Eyring & $\frac{2W\tau_{c}^{2}}{\mu_{r}}\left(\frac{B}{\tau_{B}}\right)^{2}\left[\tau_{B}\,\cosh\left(\frac{\tau_{B}}{\tau_{c}}\right)-\tau_{c}\,\sinh\left(\frac{\tau_{B}}{\tau_{c}}\right)\right]$\tabularnewline
 Carreau & $\frac{2WB^{2}I_{{\rm s,Ca}}}{\tau_{B}^{2}}$\tabularnewline
 Cross & $\frac{2WB^{2}I_{{\rm s,Cr}}}{\tau_{B}^{2}}$\tabularnewline
 Bingham & $\frac{2W}{C'}\left(\frac{B}{\tau_{B}}\right)^{2}\left[\frac{\tau_{B}^{3}}{3}-\frac{\tau_{0}\tau_{B}^{2}}{2}+\frac{\tau_{0}^{3}}{6}\right]$\tabularnewline
 Herschel-Bulkley & $\frac{2W}{\sqrt[n]{C}}\left(\frac{B}{\tau_{B}}\right)^{2}\left[\frac{n\left(n\tau_{0}+n\tau_{B}+\tau_{B}\right)\left(\tau_{B}-\tau_{0}\right)^{1+1/n}}{\left(2n^{2}+3n+1\right)}\right]$\tabularnewline
 Casson & $\frac{2W}{K}\left(\frac{B}{\tau_{B}}\right)^{2}\left[\frac{\tau_{B}^{3}}{3}-\frac{4\sqrt{\tau_{0}}\tau_{B}^{5/2}}{5}+\frac{\tau_{0}\tau_{B}^{2}}{2}-\frac{\tau_{0}^{3}}{30}\right]$\tabularnewline
 \hline
\end{tabular}
}
\end{center}
\end{table}

\renewcommand{\arraystretch}{1.2}


\clearpage
\section{Implementation and Results} \label{Implementation}

The optimization method, as described in the last section, was implemented in a computer code using
five numerical optimization algorithms: three deterministic which are Conjugate Gradient,
Quasi-Newton and Nelder-Mead \cite{PressTVF2002}; and two stochastic which are the Stochastic
Global algorithm of Boender \etal\ \cite{BoenderKTS1982}\footnote{See also:
\url{http://jblevins.org/mirror/amiller/global.txt} web page.} and a generic simulated annealing
algorithm \cite{SochiPresSA2014}. These five algorithms produce similar solutions with different
levels of accuracy and convergence rate. The sample results presented in this paper are obtained
mostly from the Stochastic Global algorithm which is overwhelmingly the most accurate and reliable
one. In addition, standard numerical integration and bisection solution techniques, as well as
standard algorithms for evaluating complicated functions like hypergeometric, were employed.

The newly proposed variational method was then employed to obtain solutions for the flow of nine
types of fluid through pipes and slits. The nine types of fluid are: Newtonian, power law, Ellis,
Ree-Eyring, Carreau, Cross, Bingham, Herschel-Bulkley and Casson. The results obtained from the new
method using wide ranges of fluid and conduit parameters were thoroughly compared to the results
obtained from the traditional methods of fluid mechanics and the former variational method.

In all the investigated cases the three methods produced very similar results within acceptable
error margins. One exception is the Ellis model for which EL has not been formulated and
implemented. Another exception is the viscoplastic fluids where the EL method differs significantly
when the yield stress value is high. This is justified by the fact that the EL method was
formulated and implemented for the non-viscoplastic fluids specifically and hence, as we
demonstrated in our previous investigations \cite{SochiVariational2013, SochiSlitPaper2014,
SochiVarNonNewt2014}, it is just an approximation for the viscoplastic fluids which is a good one
only when the yield stress value is low. However, we believe that even the EL method can be
re-formulated and re-implemented to include viscoplastic fluids, and hence it can produce similar
results to the other methods even for the high yield stress fluids, although we did not make any
effort in the current study to do so.

A representative sample of these results obtained from the three methods are presented in Figures
\ref{GP}, \ref{GS}, \ref{QP} and \ref{QS} where the fluid and conduit parameters of these examples
are given in Tables \ref{PipeFDTable} and \ref{SlitFDTable}. In these figures, the analytical
solution is represented by the solid line while the solutions from the two variational methods are
represented by the circles. The exception, as indicated earlier, is the Ellis and viscoplastic
fluids where the circles represent only the DM method. The reason for combining the solutions of
the two variational methods in a single representation is that they produce very similar results
and hence there is no point in plotting them separately.

As seen in these examples, the two variational solutions agree very well with the analytical
solutions. The minor departure in some cases between the two variational methods on one hand and
the analytical on the other is due mainly to the nature of the variational methods as they heavily
rely on numerical techniques, mainly bisection solvers and numerical integration, which is not the
case with the analytical solutions since they are evaluated directly. As indicated above, unlike
the EL approach which is a good approximation for the viscoplastic fluids only if their yield
stress is low, the DM approach produces `exact' solutions, considering the numerical errors
introduced by the heavy use of numerical techniques, even for the fluids with high yield stress.

\newcommand{\Hs}      {\hspace{-0.1cm}} %
\newcommand{\CIF}     {\centering \includegraphics[width=1.85in]} %
\newcommand{\Vmin}    {\vspace{-0.2cm}} %

\clearpage

\begin{figure}
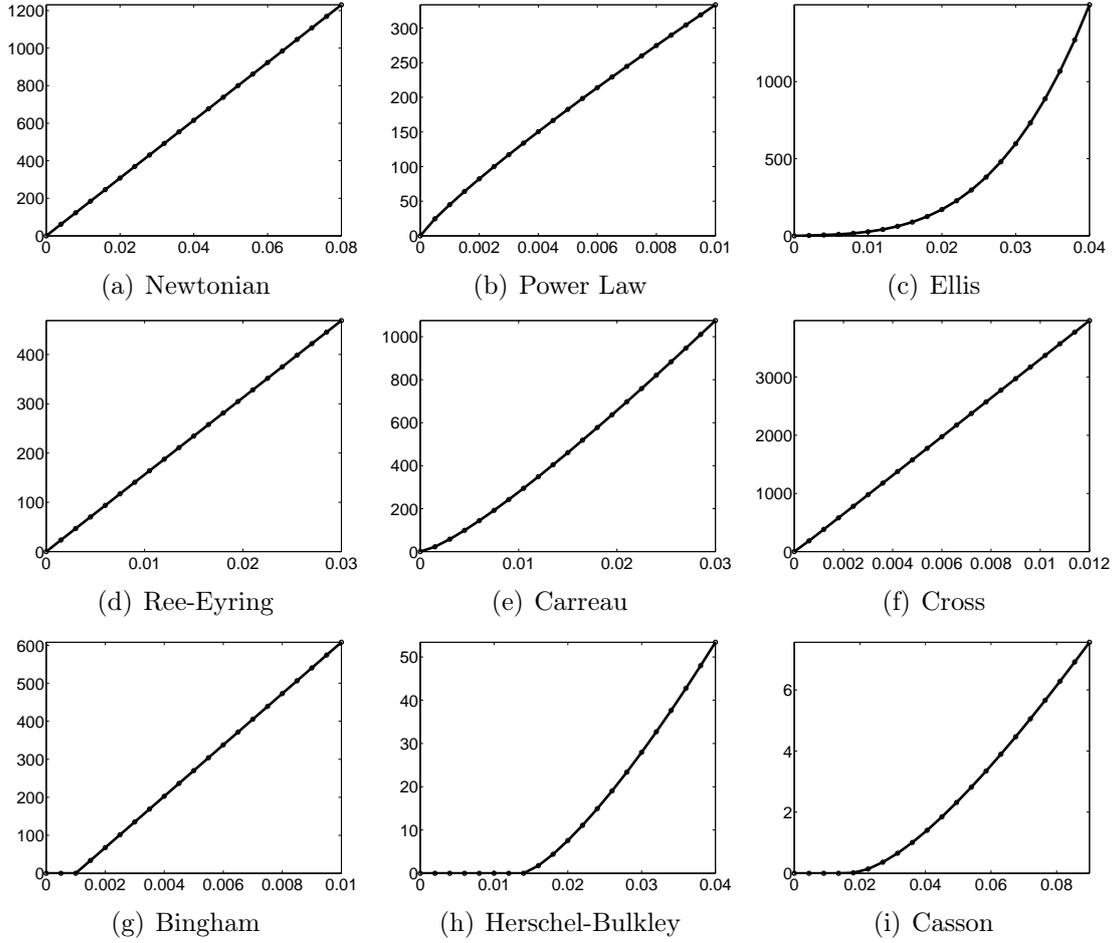

\centering %
\subfigure[Newtonian]%
{\begin{minipage}[b]{0.33\textwidth} \CIF {g/GP1}
\end{minipage}}
\Hs %
\subfigure[Power Law]%
{\begin{minipage}[b]{0.33\textwidth} \CIF {g/GP2}
\end{minipage}}
\Hs %
\subfigure[Ellis]%
{\begin{minipage}[b]{0.33\textwidth} \CIF {g/GP3}
\end{minipage}} \Vmin

%
\centering %
\subfigure[Ree-Eyring]%
{\begin{minipage}[b]{0.33\textwidth} \CIF {g/GP4}
\end{minipage}}
\Hs %
\subfigure[Carreau]%
{\begin{minipage}[b]{0.33\textwidth} \CIF {g/GP5}
\end{minipage}}
\Hs %
\subfigure[Cross]%
{\begin{minipage}[b]{0.33\textwidth} \CIF {g/GP6}
\end{minipage}}  \Vmin

%
\centering %
\subfigure[Bingham]%
{\begin{minipage}[b]{0.33\textwidth} \CIF {g/GP7}
\end{minipage}}
\Hs %
\subfigure[Herschel-Bulkley]%
{\begin{minipage}[b]{0.33\textwidth} \CIF {g/GP8}
\end{minipage}}
\Hs %
\subfigure[Casson]%
{\begin{minipage}[b]{0.33\textwidth} \CIF {g/GP9}
\end{minipage}}
\caption{Comparing the analytical solution (solid line) to the variational solutions (circles) of
$\gamma$ in s$^{-1}$ (vertical axis) versus $r$ in m (horizontal axis) for the flow of the nine
fluid models in pipes. The EL solutions are not represented for the Ellis and viscoplastic fluids.
The pipe and fluid parameters are given in Table \ref{PipeFDTable}, where in all cases $\Delta
p=500$~Pa. \label{GP}}
\end{figure}

\clearpage

\begin{figure}
\centering %
\subfigure[Newtonian]%
{\begin{minipage}[b]{0.33\textwidth} \CIF {g/GS1}
\end{minipage}}
\Hs %
\subfigure[Power Law]%
{\begin{minipage}[b]{0.33\textwidth} \CIF {g/GS2}
\end{minipage}}
\Hs %
\subfigure[Ellis]%
{\begin{minipage}[b]{0.33\textwidth} \CIF {g/GS3}
\end{minipage}} \Vmin

%
\centering %
\subfigure[Ree-Eyring]%
{\begin{minipage}[b]{0.33\textwidth} \CIF {g/GS4}
\end{minipage}}
\Hs %
\subfigure[Carreau]%
{\begin{minipage}[b]{0.33\textwidth} \CIF {g/GS5}
\end{minipage}}
\Hs %
\subfigure[Cross]%
{\begin{minipage}[b]{0.33\textwidth} \CIF {g/GS6}
\end{minipage}}  \Vmin

%
\centering %
\subfigure[Bingham]%
{\begin{minipage}[b]{0.33\textwidth} \CIF {g/GS7}
\end{minipage}}
\Hs %
\subfigure[Herschel-Bulkley]%
{\begin{minipage}[b]{0.33\textwidth} \CIF {g/GS8}
\end{minipage}}
\Hs %
\subfigure[Casson]%
{\begin{minipage}[b]{0.33\textwidth} \CIF {g/GS9}
\end{minipage}}
\caption{Comparing the analytical solution (solid line) to the variational solutions (circles) of
$\gamma$ in s$^{-1}$ (vertical axis) versus $z$ in m (horizontal axis) for the flow of the nine
fluid models in slits. The EL solutions are not represented for the Ellis and viscoplastic fluids.
The slit and fluid parameters are given in Table \ref{SlitFDTable}, where in all cases $\Delta
p=700$~Pa. \label{GS}}
\end{figure}

\clearpage

\begin{figure}
\centering %
\subfigure[Newtonian]%
{\begin{minipage}[b]{0.33\textwidth} \CIF {g/QP1}
\end{minipage}}
\Hs %
\subfigure[Power Law]%
{\begin{minipage}[b]{0.33\textwidth} \CIF {g/QP2}
\end{minipage}}
\Hs %
\subfigure[Ellis]%
{\begin{minipage}[b]{0.33\textwidth} \CIF {g/QP3}
\end{minipage}} \Vmin

%
\centering %
\subfigure[Ree-Eyring]%
{\begin{minipage}[b]{0.33\textwidth} \CIF {g/QP4}
\end{minipage}}
\Hs %
\subfigure[Carreau]%
{\begin{minipage}[b]{0.33\textwidth} \CIF {g/QP5}
\end{minipage}}
\Hs %
\subfigure[Cross]%
{\begin{minipage}[b]{0.33\textwidth} \CIF {g/QP6}
\end{minipage}}  \Vmin

%
\centering %
\subfigure[Bingham]%
{\begin{minipage}[b]{0.33\textwidth} \CIF {g/QP7}
\end{minipage}}
\Hs %
\subfigure[Herschel-Bulkley]%
{\begin{minipage}[b]{0.33\textwidth} \CIF {g/QP8}
\end{minipage}}
\Hs %
\subfigure[Casson]%
{\begin{minipage}[b]{0.33\textwidth} \CIF {g/QP9}
\end{minipage}}
\caption{Comparing the analytical solution (solid line) to the variational solutions (circles) of
$Q$ in m$^3$.s$^{-1}$ (vertical axis) versus $\Delta p$ in Pa (horizontal axis) for the flow of the
nine fluid models in pipes. The EL solutions are not represented for the Ellis and viscoplastic
fluids. The pipe and fluid parameters are given in Table \ref{PipeFDTable}. \label{QP}}
\end{figure}

\clearpage

\begin{figure}
\centering %
\subfigure[Newtonian]%
{\begin{minipage}[b]{0.33\textwidth} \CIF {g/QS1}
\end{minipage}}
\Hs %
\subfigure[Power Law]%
{\begin{minipage}[b]{0.33\textwidth} \CIF {g/QS2}
\end{minipage}}
\Hs %
\subfigure[Ellis]%
{\begin{minipage}[b]{0.33\textwidth} \CIF {g/QS3}
\end{minipage}} \Vmin

%
\centering %
\subfigure[Ree-Eyring]%
{\begin{minipage}[b]{0.33\textwidth} \CIF {g/QS4}
\end{minipage}}
\Hs %
\subfigure[Carreau]%
{\begin{minipage}[b]{0.33\textwidth} \CIF {g/QS5}
\end{minipage}}
\Hs %
\subfigure[Cross]%
{\begin{minipage}[b]{0.33\textwidth} \CIF {g/QS6}
\end{minipage}}  \Vmin

%
\centering %
\subfigure[Bingham]%
{\begin{minipage}[b]{0.33\textwidth} \CIF {g/QS7}
\end{minipage}}
\Hs %
\subfigure[Herschel-Bulkley]%
{\begin{minipage}[b]{0.33\textwidth} \CIF {g/QS8}
\end{minipage}}
\Hs %
\subfigure[Casson]%
{\begin{minipage}[b]{0.33\textwidth} \CIF {g/QS9}
\end{minipage}}
\caption{Comparing the analytical solution (solid line) to the variational solutions (circles) of
$Q$ in m$^3$.s$^{-1}$ (vertical axis) versus $\Delta p$ in Pa (horizontal axis) for the flow of the
nine fluid models in slits. The EL solutions are not represented for the Ellis and viscoplastic
fluids. The slit and fluid parameters are given in Table \ref{SlitFDTable}. \label{QS}}
\end{figure}

\clearpage

\begin{table} [!h]
\caption{Fluid and pipe parameters for the examples of Figures \ref{GP} and \ref{QP}. SI units
apply to all dimensional quantities as given in Nomenclature \S\ \ref{Nomenclature}.
\label{PipeFDTable}}
\begin{center} 
{
\begin{tabular}{|l|l|c|c|}
\hline
    Model & Fluid Properties & $R$ & $L$\tabularnewline
    \hline
    Newtonian & $\mu_{o}=0.025$ & 0.08 & 0.65\tabularnewline
    \hline
    Power Law & $k=0.033,\,\,\,\,\,\,\,\,\,\, n=1.15$ & 0.01 & 0.095\tabularnewline
    \hline
    Ellis & $\mu_{e}=1.42,\,\,\,\,\,\,\,\,\,\,\tau_{h}=15,\,\,\,\,\,\,\,\,\,\,\alpha=3.3$ & 0.04 & 0.15\tabularnewline
    \hline
    Ree-Eyring & $\mu_{r}=0.02,\,\,\,\,\,\,\,\,\,\,\tau_{c}=300$ & 0.03 & 0.8\tabularnewline
    \hline
    Carreau & $\mu_{0}=0.1,\,\,\,\,\,\,\,\,\,\,\mu_{i}=0.008,\,\,\,\,\,\,\,\,\,\,\lambda=1.2,\,\,\,\,\,\,\,\,\,\, n=0.65$ & 0.03 & 0.45\tabularnewline
    \hline
    Cross & $\mu_{0}=0.15,\,\,\,\,\,\,\,\,\,\,\mu_{i}=0.005,\,\,\,\,\,\,\,\,\,\,\lambda=7.9,\,\,\,\,\,\,\,\,\,\, m=0.8$ & 0.012 & 0.15\tabularnewline
    \hline
    Bingham & $C'=0.037,\,\,\,\,\,\,\,\,\,\,\tau_{0}=2.5$ & 0.01 & 0.1\tabularnewline
    \hline
    Herschel-Bulkley & $C=0.47,\,\,\,\,\,\,\,\,\,\,\tau_{0}=5.0,\,\,\,\,\,\,\,\,\,\, n=0.75$ & 0.04 & 0.7\tabularnewline
    \hline
    Casson & $K=0.75,\,\,\,\,\,\,\,\,\,\,\tau_{0}=3.0$ & 0.09 & 1.33\tabularnewline
    \hline
\end{tabular}
}
\end{center}
\end{table}


\begin{table} [!h]
\caption{Fluid and slit parameters for the examples of Figures \ref{GS} and \ref{QS}, where in all
cases $W=1.0$~m. SI units apply to all dimensional quantities as given in Nomenclature \S\
\ref{Nomenclature}. \label{SlitFDTable}}
\begin{center} 
{
\begin{tabular}{|l|l|c|c|}
    \hline
    Model & Fluid Properties & $B$ & $L$\tabularnewline
    \hline
    Newtonian & $\mu_{o}=0.13$ & 0.02 & 1.2\tabularnewline
    \hline
    Power Law & $k=0.073,\,\,\,\,\,\,\,\,\,\, n=0.67$ & 0.01 & 0.95\tabularnewline
    \hline
    Ellis & $\mu_{e}=0.049,\,\,\,\,\,\,\,\,\,\,\tau_{h}=5.0,\,\,\,\,\,\,\,\,\,\,\alpha=2.9$ & 0.011 & 1.95\tabularnewline
    \hline
    Ree-Eyring & $\mu_{r}=0.57,\,\,\,\,\,\,\,\,\,\,\tau_{c}=75$ & 0.05 & 11.5\tabularnewline
    \hline
    Carreau & $\mu_{0}=0.32,\,\,\,\,\,\,\,\,\,\,\mu_{i}=0.096,\,\,\,\,\,\,\,\,\,\,\lambda=0.75,\,\,\,\,\,\,\,\,\,\, n=0.85$ & 0.023 & 0.75\tabularnewline
    \hline
    Cross & $\mu_{0}=0.015,\,\,\,\,\,\,\,\,\,\,\mu_{i}=0.007,\,\,\,\,\,\,\,\,\,\,\lambda=2.56,\,\,\,\,\,\,\,\,\,\, m=0.73$ & 0.035 & 2.13\tabularnewline
    \hline
    Bingham & $C'=0.48,\,\,\,\,\,\,\,\,\,\,\tau_{0}=4.3$ & 0.014 & 1.03\tabularnewline
    \hline
    Herschel-Bulkley & $C=0.03,\,\,\,\,\,\,\,\,\,\,\tau_{0}=5.2,\,\,\,\,\,\,\,\,\,\, n=1.45$ & 0.035 & 3.09\tabularnewline
    \hline
    Casson & $K=0.12,\,\,\,\,\,\,\,\,\,\,\tau_{0}=2.15$ & 0.017 & 2.33\tabularnewline
    \hline
\end{tabular}
}
\end{center}
\end{table}


\clearpage
\section{Conclusions} \label{Conclusions}

In this paper we presented a variational approach for finding the flow solutions in one dimensional
flow that applies easily to circular pipes and plane slits. The method, which is demonstrated using
nine types of fluid, can be employed to obtain all the required flow parameters which include shear
stress, local viscosity, shear rate, speed profile and volumetric flow rate. We also presented,
through the application of the Dirichlet principle, a theoretical justification for the application
of minimizing the total stress profile as a basis for our variational approaches including the EL
method.

Thorough comparisons were made both to the analytical solutions obtained from the traditional
methods of fluid dynamics and to the analytical or semi analytical solutions obtained from the EL
method. In all cases, the three methods produced very close results where the differences can be
explained by the numerical errors introduced by heavy use of numerical methods like numerical
integration, bisection solvers and numeric evaluation of complicated functions. The exception is
the viscoplastic fluids for which the EL method cannot provide reliable solutions when the yield
stress is high due to the particular formulation and implementation of this method which is
restricted to non-viscoplastic fluids. The EL method also has not been formulated and implemented
in the past for the Ellis fluid and we did not make any effort to do so in the current study.

Apart from being useful on its own for resolving the flow field in the given conduits and obtaining
all the required parameters, the proposed DM method adds more support to the previous EL approach,
which is based on applying the Euler-Lagrange variational principle, as it confirms the results
obtained from the EL method and provides a theoretical foundation for it. Although the new method
may not be conceptually identical to the previous one, it should still lend support to the previous
one not only because the two differently-formulated variational methods produce similar results but
also because they are both based on similar variational principles. From a procedural point of
view, the two methods are equivalent because what is done in DM numerically is done in EL, as
presented in our previous studies, either analytically or partly analytically and partly
numerically. It has also been shown that the theoretical foundation of the DM method, represented
by the Dirichlet principle, endorses the particular formulation of stress minimization which EL
rests upon.

There is an obvious theoretical value of the new DM variational method which is more important than
its practical value which may be insignificant for the investigated fluid types and conduit
geometries due to the availability of the presented analytical and numerical flow solutions from
other methods. If the proposed variational principle enjoys a general applicability, which is yet
to be established beyond the one dimensional flow through pipes and slits, the method may also have
a significant practical value for the flow systems which are more complicated than the flow systems
in pipes and slits where the DM variational method may provide solutions that other methods might
fail to provide, or the DM method may require less effort to obtain the solutions than the effort
required by the other methods.

\clearpage
\section{Nomenclature}\label{Nomenclature}

\begin{supertabular}{ll}
$\alpha$                &   indicial parameter in Ellis model \\
$\gamma$                &   rate of shear strain (s$^{-1}$) \\
$\gamma_B$              &   rate of shear strain at slit wall (s$^{-1}$) \\
$\gamma_R$              &   rate of shear strain at pipe wall (s$^{-1}$) \\
$\gamma_w$              &   rate of shear strain at conduit wall (s$^{-1}$) \\
$\delta$                &   $\mu_{0}-\mu_{i}$ (Pa.s) \\
$\lambda$               &   characteristic time constant in Carreau and Cross models (s) \\
$\mu$                   &   fluid shear viscosity (Pa.s) \\
$\mu_{0}$               &   zero-shear viscosity in Carreau and Cross models (Pa.s) \\
$\mu_e$                 &   low-shear viscosity in Ellis model (Pa.s) \\
$\mu_{i}$               &   infinite-shear viscosity in Carreau and Cross models (Pa.s) \\
$\mu_{o}$               &   Newtonian viscosity (Pa.s) \\
$\mu_{r}$               &   characteristic viscosity in Ree-Eyring model (Pa.s) \\
$\tau$                  &   shear stress (Pa) \\
$\tau_0$                &   yield stress in Bingham, Herschel-Bulkley and Casson models (Pa) \\
$\tau_B$                &   shear stress at slit wall (Pa) \\
$\tau_c$                &   characteristic shear stress in Ree-Eyring model (Pa) \\
$\tau_{h}$              &   shear stress when viscosity equals $\frac{\mu_e}{2}$ in Ellis model (Pa) \\
$\tau_m$                &   shear stress at conduit center (Pa) \\
$\tau_R$                &   shear stress at pipe wall (Pa) \\
$\tau_w$                &   shear stress at conduit wall (Pa) \\
$\Omega$                &   spatial domain in Dirichlet functional (m) \\
\\
$B$                     &   slit half thickness (m) \\
$C$                     &   viscosity coefficient in Herschel-Bulkley model (Pa.s$^{n}$) \\
$C'$                    &   viscosity coefficient in Bingham model (Pa.s) \\
$D$                     &   Dirichlet functional (Pa$^2$.m$^{-1}$) \\
$f$                     &   $\lambda^{m}\gamma_{w}^{m}$ \\
$_{2}F_{1}$             &   hypergeometric function \\
$g$                     &   $1+f$ \\
$G$                     &   constant in the reduced momentum equation (Pa.m$^{-1}$) \\
$I_{\rm p,Ca}$          &   definite integral for Carreau model pipe flow (Pa$^3$.s$^{-1}$) \\
$I_{\rm p,Cr}$          &   definite integral for Cross model pipe flow (Pa$^3$.s$^{-1}$) \\
$I_{\rm s,Ca}$          &   definite integral for Carreau model slit flow (Pa$^2$.s$^{-1}$) \\
$I_{\rm s,Cr}$          &   definite integral for Cross model slit flow (Pa$^2$.s$^{-1}$) \\
$k$                     &   viscosity coefficient in power law model (Pa.s$^n$) \\
$K$                     &   viscosity coefficient in Casson model (Pa.s) \\
$L$                     &   conduit length (m) \\
$m$                     &   indicial parameter in Cross model \\
$n$                     &   flow behavior index in power law, Carreau and Herschel-Bulkley models \\
$\Delta p$              &   pressure drop across conduit length (Pa) \\
$Q$                     &   volumetric flow rate (m$^{3}$.s$^{-1}$) \\
$r$                     &   radius (m) \\
$R$                     &   pipe radius (m) \\
$s$                     &   spatial coordinate representing $r$ for pipe and $z$ for slit (m) \\
$T$                     &   stress domain in Dirichlet functional (Pa) \\
$v$                     &   fluid speed in the flow direction (m.s$^{-1}$) \\
$W$                     &   slit width (m) \\
$z$                     &   spatial coordinate of slit thickness (m) \\
\\
DM                      &   variational method based on applying Dirichlet principle \\
EL                      &   variational method based on applying Euler-Lagrange principle \\
\end{supertabular}

\clearpage
\phantomsection \addcontentsline{toc}{section}{References} %
\bibliographystyle{unsrt}

\end{document}